\documentclass[aps,preprint,nofootinbib,superscriptaddress,prc]{revtex4}
\usepackage{amssymb}

\usepackage{epsfig}
\usepackage[bookmarksnumbered,bookmarksopen,colorlinks,citecolor=blue,linkcolor=blue]{hyperref}
\begin{document}


\title{Further improvements on a global nuclear mass model}

\author{Min Liu}
\affiliation{Department of Physics, Guangxi Normal University,
Guilin 541004, P. R. China}

\author{Ning Wang }
\thanks{Corresponding author: wangning@gxnu.edu.cn}
\affiliation{Department of Physics, Guangxi Normal University,
Guilin 541004, P. R. China}

\author{Yangge Deng }
\affiliation{Department of Physics, Guangxi Normal University,
Guilin 541004, P. R. China}

\author{Xizhen Wu}
\affiliation{China Institute of Atomic Energy, Beijing 102413, P.
R. China}

\begin{abstract}
The semi-empirical macroscopic-microscopic mass formula is further
improved by considering some residual corrections. The rms
deviation from 2149 known nuclear masses is significantly reduced
to 336 keV, even lower than that achieved with the best of the
Duflo-Zuker models. The $\alpha$-decay energies of super-heavy
nuclei, the Garvey-Kelson relations and the isobaric multiplet
mass equation (IMME) can be reproduced remarkably well with the
model, and the predictive power of the mass model is good.  With a
systematic study of 17 global nuclear mass models, we find that
the quadratic form of the IMME is closely related to the accuracy
of nuclear mass calculations when the Garvey-Kelson relations are
reproduced reasonably well. Fulfilling both the IMME and the
Garvey-Kelson relations seems to be two necessary conditions to
improve the quality of the model prediction. Furthermore, the
$\alpha$-decay energies of super-heavy nuclei should be used as an
additional constraint on the global nuclear mass models.
\end{abstract}

\maketitle

\begin{center}
\textbf{I. INTRODUCTION}
\end{center}

The precise calculation for nuclear masses is of great importance
for not only nuclear physics but also nuclear astrophysics. In
nuclear physics, the synthesis of super-heavy nuclei
\cite{Ogan10,Naza,Sob} and nuclear symmetry energy \cite{LiBA,
Liu} attracted much attention in recent years. Accurate
predictions for the shell corrections and $\alpha$-decay energies
of super-heavy nuclei are urgently required for the synthesis of
new super-heavy nuclei. Simultaneously, nuclear masses include
important information on nuclear symmetry energy \cite{Nik,WL}
which closely relates to the structure of nuclei and the dynamical
process of nuclear reactions. In addition, the r-process
abundances in nuclear astrophysics depend strongly on the nuclear
masses of extremely neutron-rich nuclei and the shell-structure
data of neutron magic nuclei. Some nuclear mass models have been
developed with rms deviations of about several hundreds to a
thousand keV with respect to all known nuclear masses. The most
successful global mass models include: 1) various
macroscopic-microscopic mass models such as the finite range
droplet model (FRDM) \cite{Moll95}, the extended
Bethe-Weizs\"{a}cker (BW2) formula \cite{BW} and the
Weizs\"acker-Skyrme (WS) mass models proposed very recently by
Wang et al. \cite{Wang,Wang10}; 2) various microscopic mass models
based on the mean-field concept such as the non-relativistic
Hartree-Fock-Bogoliubov (HFB) approach with the Skyrme
energy-density functional \cite{HFB17} or the Gogney forces and
the relativistic mean-field (RMF) model \cite{RMF}; and 3) the
Duflo-Zuker (DZ) mass models \cite{DZ10,DZ28} which are successful
for describing known masses, with an accuracy at the level of 360
keV.  For known masses, the predictions from these global mass
models are close to each other. However, for neutron drip line
nuclei and super-heavy nuclei, the differences of the calculated
masses from these models are quite large. It is therefore
necessary to perform a systematic study of these models to check
the reliability for extrapolations.

Besides the global mass models, some local relations and equations
such as the isobaric multiplet mass equation (IMME)
\cite{Ormand,Lenzi}, the Garvey-Kelson (GK) relations
\cite{GK,GK100} and the residual proton-neutron interactions
\cite{Zhao,Zhao1} are used to analyze the isospin-symmetry
breaking effects and the consistency of nuclear mass predictions.
It is thought that the binding energies of extremely proton-rich
nuclei can be predicted at the level of $100\sim 200$ keV based on
the IMME \cite{Ormand}. As the IMME is a basic prediction leading
from the isospin concept, testing the validity of this equation is
of fundamental importance. On the other hand, it is shown that the
GK relations which are obtained from an independent-particle
picture and the charge-symmetry of nuclear force, are satisfied to
a very high level of accuracy by known masses, with an rms
deviation of about 100 keV \cite{GK100}. It is of crucial
importance to incorporate these local relations and equations in
the global nuclear mass models for exploring the missing physics
and the constraints in the models.

In our previous works \cite{Wang,Wang10}, we proposed a
semi-empirical nuclear mass formula based on the
macroscopic-microscopic method together with the Skyrme
energy-density functional, with an rms deviation of 441 keV for
2149 known masses. The isospin and mass dependence of the model
parameters such as the symmetry energy coefficient and the
constraint between mirror nuclei from the isospin symmetry in
nuclear physics play an important role for improving the accuracy
of the mass calculation. In this work, the formula will be further
improved by considering some residual corrections of nuclei. With
some empirical considerations for these residual corrections, we
find both the precision and the consistency of the mass formula
can be significantly improved. The paper is organized as follows:
In Sec. II, the mass formula and some corrections are briefly
introduced. In Sec. III, calculated nuclear masses from the
proposed model are presented and compared to the results from
other mass models. In Sec. IV, a number of global nuclear mass
models are systematically tested by using the Garvey-Kelson
relations and the IMME. Finally, a summary is given in Sec. V.

\begin{center}
\textbf{II. THE NUCLEAR MASS MODEL AND SOME CORRECTIONS}
\end{center}

In this mass formula, the total energy of a nucleus is written as
a sum of the liquid-drop energy, the Strutinsky shell correction
$\Delta E$ and the residual correction $\Delta_{\rm res}$,
\begin{eqnarray}
E (A,Z,\beta)=E_{\rm LD}(A,Z) \prod_{k \ge 2} \left (1+b_k
\beta_k^2 \right )+\Delta E (A,Z,\beta) + \Delta_{\rm res}.
\end{eqnarray}
The liquid-drop energy of a spherical nucleus $E_{\rm LD}(A,Z)$ is
described by a modified Bethe-Weizs\"acker mass formula,
\begin{eqnarray}
E_{\rm LD}(A,Z)=a_{v} A + a_{s} A^{2/3}+ E_C + a_{\rm sym} I^2 A +
a_{\rm pair}  A^{-1/3}\delta_{np} + \Delta_W
\end{eqnarray}
with the isospin asymmetry $I=(N-Z)/A$, the Coulomb energy,
\begin{eqnarray}
E_C=a_c \frac{Z^2}{A^{1/3}} \left ( 1- 0.76 Z^{-2/3} \right)
\end{eqnarray}
and the symmetry energy coefficient,
\begin{eqnarray}
 a_{\rm sym}=c_{\rm sym}\left [1-\frac{\kappa}{A^{1/3}}+ \xi  \frac{2-|I|}{ 2+|I|A}  \right
 ].
\end{eqnarray}
The $a_{\rm pair}$ term empirically describes the pairing effect
(see Ref.\cite{Wang} for details). The terms with $b_k$ describe
the contribution of nuclear deformation (including $\beta_2$,
$\beta_4$ and $\beta_6$) to the macroscopic energy. Mass
dependence of the curvatures $b_k$ in Eq.(1) is written as
\cite{Wang},
\begin{eqnarray}
b_k=\left ( \frac{k}{2} \right ) g_1A^{1/3}+\left ( \frac{k}{2}
\right )^2 g_2 A^{-1/3},
\end{eqnarray}
according to the Skyrme energy-density functional, which greatly
reduces the computation time for the calculation of deformed
nuclei.

\begin{figure}
\includegraphics[angle=-0,width= 0.8\textwidth]{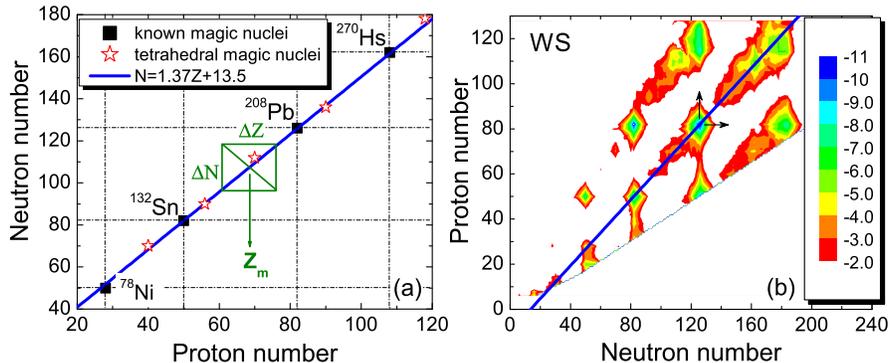}
 \caption{(Color online) (a) Positions of some known doubly-magic nuclei (squares). The stars denote the predicted tetrahedral doubly-magic
 nuclei in Refs.\cite{Cur,Maz}. (b) Shell corrections of nuclei with WS \cite{Wang}. The solid line shows the position of $N=1.37Z+13.5$.
}
\end{figure}

In addition, we propose an empirical correction term $\Delta_W$ in
$E_{\rm LD}$ to consider the Wigner-like effect of heavy nuclei
\cite{Wig} which is due to the approximate symmetry between
valance-proton and valance-neutron. In \cite{Wang,Wang10}, we
found that some heavy doubly-magic nuclei surprisingly lie along a
straight line $N=1.37Z+13.5$ (see Fig.1), which is called shell
stability line of heavy nuclei. Furthermore, some predicted
tetrahedral doubly-magic nuclei \cite{Cur,Maz} [stars in Fig.1(a)]
also lie near the line. The shell corrections of nuclei obtained
with WS \cite{Wang} are shown in Fig.1(b), and they are
approximately symmetric along the shell stability line, i.e. the
shell correction of doubly-magic nucleus $^{164}$Pb is close to
that of $^{176}$Sn. In other words, the change of the shell energy
by adding several neutrons to the doubly-magic nucleus $^{208}$Pb
is close to that by adding several protons to $^{208}$Pb because
of the charge-independence of nuclear force and the fact that the
Coulomb interaction is weak relative to the strong interaction. To
consider the symmetry and interaction between valance-proton and
valance-neutron, we phenomenologically write the Wigner-like term
as,
\begin{eqnarray}
\Delta_W = c_{\rm w}\left (e^{|I|}-e^{-|\eta|} \right ).
\end{eqnarray}
Where, $\eta=\frac{\Delta Z \; \Delta N}{Z_{\rm m}}$ denotes the
effective distance of a nucleus to the shell stability line. The
values of $\Delta Z$, $\Delta N$ and $Z_{\rm m}$ are uniquely
determined by a rectangle [see the green rectangle in Fig.1(a)]
based on the position of a nucleus under consideration and the
shell stability line. Similar to the known Wigner effect causing a
cusp for light nuclei near the $N=Z$ line, the Wigner-like term
$\Delta_W$ causes a cusp for heavy nuclei limited to the close
vicinity of the shell stability line,  which reduces the rms
deviation with respect to masses by about $5\%$.

The microscopic shell correction
\begin{eqnarray}
\Delta E=c_1 E_{\rm sh} + |I| E_{\rm sh}^{\prime}
\end{eqnarray}
is obtained with the traditional Strutinsky procedure \cite{Strut}
by setting the order $p=6$ of the Gauss-Hermite polynomials and
the smoothing parameter $\gamma=1.2\hbar\omega_0$ with
$\hbar\omega_0=41 A^{-1/3}$ MeV. $E_{\rm sh}$ and $E_{\rm
sh}^{\prime}$ denote the shell energy of a nucleus and of its
mirror nucleus, respectively. The $|I|$ term in $\Delta E$ is to
take into account the mirror nuclei constraint \cite{Wang10} from
the isospin symmetry, with which the accuracy of the mass model
can be significantly improved (by $10\%$). The single-particle
levels are obtained under an axially deformed Woods-Saxon
potential \cite{Cwoik}. Simultaneously, the isospin-dependent
spin-orbit strength is adopted based on the Skyrme energy-density
functional,
\begin{eqnarray}
\lambda= \lambda_0 \left ( 1+\frac{N_i}{A} \right )
\end{eqnarray}
with $N_i=Z$ for protons and $N_i=N$ for neutrons, which strongly
affects the shell structure of neutron-rich nuclei and super-heavy
nuclei.

Some other residual corrections caused by the microscopic shell
effect are empirically written as a sum of three terms,
\begin{eqnarray}
\Delta_{\rm res}=\Delta_M +  \Delta_{P}+ \Delta_T.
\end{eqnarray}
The first term $\Delta_{M}$ is to further consider mirror nuclei
effect. In our previous work \cite{Wang10}, we found that the
shell correction difference $|\Delta E-\Delta E^\prime|$ between a
nucleus and its mirror nucleus is small due to the isospin
symmetry in nuclear physics. Here, we empirically write
\begin{eqnarray}
\Delta_M=   c_2 (S+S_1)(1-I^2 A),
\end{eqnarray}
with
\begin{eqnarray}
S=\frac{(\Delta E - \Delta E^\prime)^2}{T(T+1)}.
\end{eqnarray}
Here, $T=\left
 |\frac{N-Z}{2} \right | $ denotes the isospin of a nucleus, and we set $S=0$ for nuclei with $N=Z$.
$S_1$ is the corresponding value of a neighboring isobaric nucleus
$(A,Z_1)$ with $Z_1=Z+1$ for $N<Z$ cases and $Z-1$ for $N\ge Z$
cases. The $\Delta_{M}$  term effectively describes the residual
mirror nuclei and isospin effect in nuclei and significantly
improves the accuracy of mass calculation by about 50 keV.

The second term $\Delta_P$ further considers the residual pairing
corrections of nuclei which may  be phenomenologically given by
the second differences of the masses. That is, in addition to the
$a_{\rm pair}$ term in $E_{\rm LD}$ for describing the pairing
correction, we further consider the term
\begin{eqnarray}
 \Delta_P =
\frac{1}{2}\frac{\partial^2 E_{\rm sh}}{\partial A^2} {\Big |}_T
\end{eqnarray}
with a step size of $\Delta A=2$ in the calculations. This term
improves the smoothness of the mass surface assumed in the
Garvey-Kelson relations \cite{GK}, and simultaneously reduces the
rms deviation by about $4\%$ and $6\%$ with respect to the masses
and the neutron separation energies, respectively. In addition,
for describing the influence of triaxial (or tetrahedral)
deformation \cite{Cur,Maz} effectively, we proposed a
parameterized formula in our previous work \cite{Wang10}. Here the
formula is slightly extended to include the contribution of the
protons, $\Delta_T = -0.7 \left [\cos\left (2\pi
\frac{N}{16}\right )+\cos \left (2 \pi \frac{N}{20} \right ) +\cos
\left (2 \pi \frac{Z}{52} \right ) \right] A^{-1/3}$ MeV.

\begin{table}
\caption{ Model parameters of the mass formula WS3. }
\begin{tabular}{cccc }
\hline\hline
  Quantity                         & ~~~~~~Value~~~~~~& ~~~ Quantity~~~       & ~~~Value~~~\\ \hline
 $a_v  \; $ (MeV)                  &   $-15.5485$   & $g_1 $                            &   0.01037 \\
 $a_s \; $  (MeV)                  &   17.4663      &  $g_2 $                           &   $-0.5071$ \\
 $a_c \; $ (MeV)                   &   0.7128       &  $V_0$ (MeV)                      &   $-45.2092$ \\
 $c_{\rm sym} $(MeV)               &   29.1174      &   $r_0$ (fm)                      &   1.3848   \\
 $\kappa \;  $                     &   1.3437       &   $a $ (fm)                       &   0.7617   \\
 $\xi \;  $                        &   1.1865       &  $\lambda_0$                      &   26.6744  \\
  $a_{\rm pair} $(MeV)             &   $-6.2299$    &  $c_1  \; $                       &   0.6454   \\
  $c_{\rm w} $ (MeV)               &   1.0490       &  $c_2  \; ({\rm MeV} ^{-1})$      &   1.6179  \\
 \hline\hline
\end{tabular}
\end{table}

\begin{table}
\caption{ rms $\sigma$ deviations between data AME2003 \cite{Audi}
and predictions of some models (in keV). The line $\sigma (M)$
refers to all the 2149 measured masses, the line $\sigma (S_n)$ to
the 1988 measured neutron separation energies $S_n$, the line
$\sigma (Q_\alpha)$ to the $\alpha$-decay energies of 46
super-heavy nuclei. $\sigma ({\rm GK})$ presents the rms deviation
with respect to the GK mass estimates. $\sigma (b)$ denotes the
rms deviation from the formula \cite{Ormand} $b_{\rm
fit}=0.710A^{2/3}-0.946$ for the $b$ coefficients of 179 neutron
drip line nuclei. }
\begin{tabular}{cccccccccccc}
 \hline\hline
       & ETF2\cite{Lium} & BW2\cite{BW} & FRDM\cite{Moll95}  & HFB-17\cite{HFB17} & DZ10\cite{DZ10} & WS\cite{Wang} & WS*\cite{Wang10}& DZ31\cite{DZ10} & DZ28\cite{DZ28}  & WS3 \\
\hline
 $\sigma  (M)$         & $3789$ &$1594$ & $656$  & $581$ & $561$ & $516$ & $441$ & $362$ & $360$ &  $336 $\\
 $\sigma  (S_n)$       & $1300$ & $586$ & $399$  & $506$ & $342$ & $346$ & $332$ & $299$ & $306$ &  $286 $\\
 $\sigma  (Q_\alpha)$  & $557$ & $1233$ & $566$  & $-$   & $916$ & $284$ & $263$ & $1052$& $936$ &  $248 $\\
 $\sigma  ({\rm GK})$  & $24$  & $129$  & $337$  & $496$ & $100$ & $151$ & $165$ & $115$ & $133$ &  $131 $\\
 $\sigma  (b)$         & $2734$& $759$  & $-$    & $-$   & $870$ & $283$ & $450$ & $630$ & $-$   &  $449 $\\

 \hline\hline
\end{tabular}
\end{table}

\begin{figure}
\includegraphics[angle=-0,width= 0.65\textwidth]{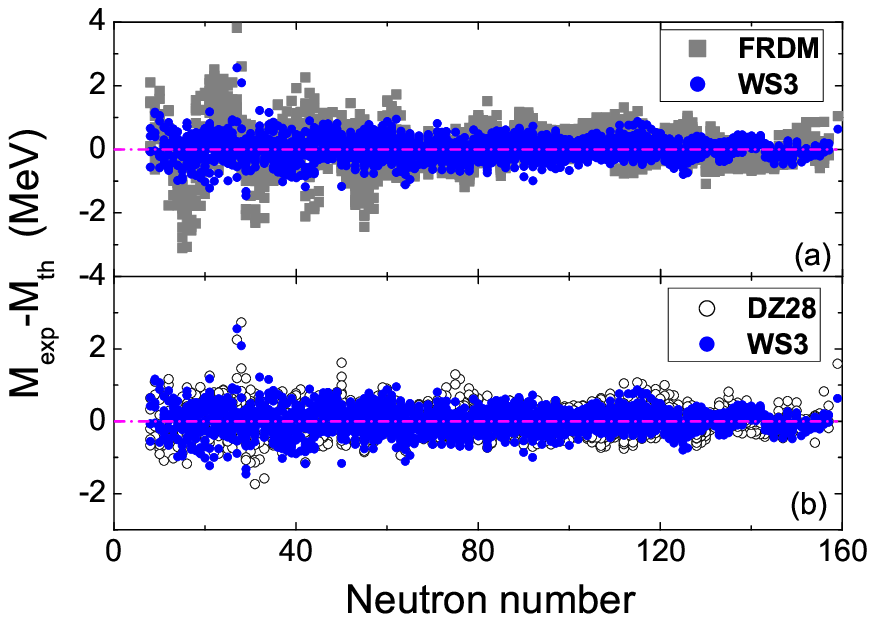}
 \caption{(Color online) (a) Deviations of calculated nuclear masses from the experimental data as a function of neutron
 number. The squares and the solid circles denote the results of FRDM and  WS3, respectively. (b) The same as (a), but with the results of DZ28 model for comparison. }
\end{figure}

\begin{center}
\textbf{III. CALCULATED MASSES FROM THE MODEL}
\end{center}

Based on the 2149 ($N$ and $Z\ge8$) measured nuclear masses
AME2003 \cite{Audi}, we obtain the optimal model parameters which
is labelled as WS3 and listed in Table I. We show in Table II the
rms deviations $\sigma (M)$ between the 2149 experimental masses
and predictions of some models (in keV). $\sigma (S_n)$ denotes
the rms deviation to the 1988 measured neutron separation energies
$S_n$. $\sigma (Q_\alpha)$ in Table II denotes the rms deviation
with respect to the $\alpha$-decay energies of 46 super-heavy
nuclei ($Z\ge 106$) \cite{Wang10}. $\sigma ({\rm GK})$ presents
the rms deviation with respect to the GK mass estimates and
$\sigma (b)$ denotes the rms deviation from the formula
\cite{Ormand} $b_{\rm fit}=0.710A^{2/3}-0.946$ for the $b$
coefficients of 179 neutron drip line nuclei, which will be
discussed in the next section. Here, ETF2 denotes the extended
Thomas-Fermi approach including all terms up to second order in
the spatial derivatives \cite{Lium} together with the Skyrme force
SkP \cite{SkP}, in which the deformations and shell corrections of
nuclei are not involved. BW2 means the extended
Bethe-Weizs\"{a}cker formula \cite{BW} in which the shell
corrections of nuclei are described as a function of
valence-nucleon number assuming the magic numbers being
$2,8,20,28,50,82,126,184$. FRDM and HFB-17 denote the finite-range
droplet model \cite{Moll95} and the latest Hartree-Fock-Bogoliubov
(HFB) model with the improved Skyrme energy-density functional
\cite{HFB17}, respectively. DZ10, DZ28 and DZ31 denote the
Duflo-Zuker mass models with 10, 28 and 31 parameters,
respectively \cite{DZ10,DZ28}.

The rms deviation from the 2149 masses with WS3 is remarkably
reduced to 336 keV, much smaller than the results from the finite
range droplet model (FRDM)  and the latest HFB-17 calculations,
even lower than that achieved with the best of the Duflo-Zuker
models. In Fig. 2, we show the deviations from the experimental
masses as a function of neutron number. The squares and the solid
circles in Fig.2(a) denote the results from the FRDM and WS3,
respectively. The accuracy of mass calculation especially for
light nuclei is significantly improved in the WS3 model. The open
circles in Fig.2(b) denote the results from the DZ28 model.
Compared to the DZ28 model, the rms deviations from masses and
neutron separation energies are reduced by $7\%$. To further test
the extrapolation of our mass model, we also calculate the average
deviation from four very recently measured masses for $^{63}$Ge,
$^{65}$As, $^{67}$Se, and $^{71}$Kr \cite{Xu}.  Our result is 118
keV, much smaller than the results from DZ28 (321 keV) and FRDM
(680 keV).

\begin{figure}
\includegraphics[angle=-0,width= 0.7\textwidth]{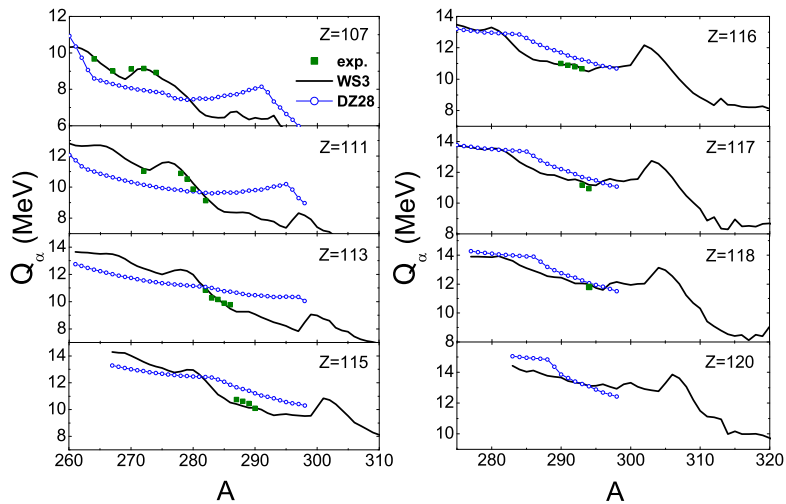}
 \caption{(Color online) $\alpha$-decay energies of some super-heavy nuclei calculated with WS3 (solid curve) and DZ28 (open circles).
 The squares denote the experimental data taken from \cite{Wang10} (and references therein).}
\end{figure}

\begin{figure}
\includegraphics[angle=-0,width= 0.8 \textwidth]{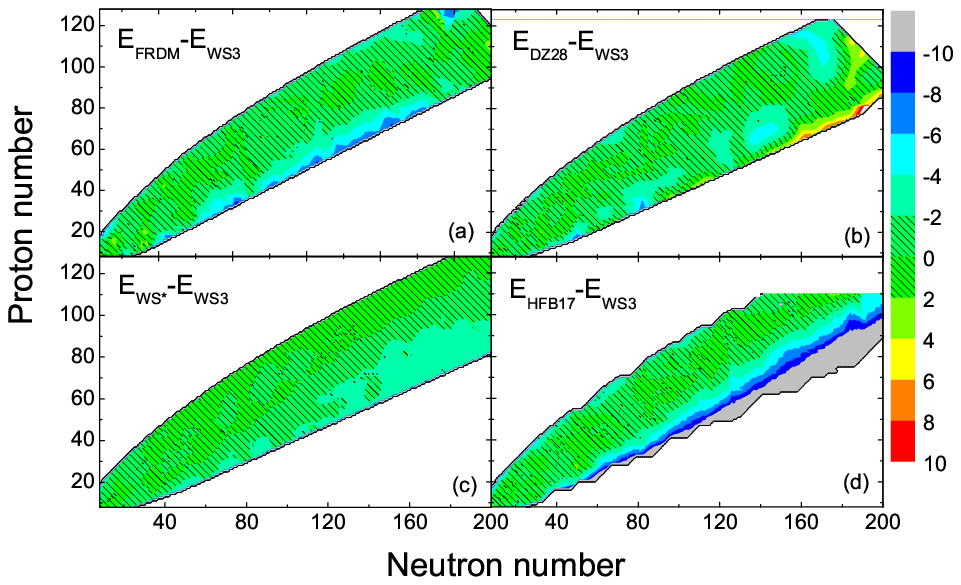}
 \caption{(Color online) Deviations of the calculated nuclear total energies at ground state in this work from the results of (a) FRDM, (b) DZ28, (c) WS* and (d) HFB-17, respectively. The
calculated masses with FRDM, DZ28, WS* and HFB-17 are taken from
\cite{Moll95}, \cite{DZ28}, \cite{Wang10} and \cite{HFB17},
respectively. The shades denote the region with deviations smaller
than 2 MeV.}
\end{figure}

It is known that the $\alpha$-decay energies of super-heavy nuclei
\cite{Ogan10} have been measured with a high precision, which
provides us with useful data for testing our mass model. The rms
deviation $\sigma (Q_\alpha)$ with respect to the $\alpha$-decay
energies of 46 super-heavy nuclei  from the WS3 model is
remarkably reduced to 248 keV. The difference between the
calculated $\alpha$-decay energies of the super-heavy nuclei and
the measured data is within $\Delta Q_\alpha=\pm 0.4$ MeV for
odd-$Z$ nuclei. The corresponding results from Sobiczewski
\cite{Ogan11} and FRDM are $ \pm 0.8$ and $ \pm 1.0$ MeV,
respectively. Here, we would like to emphasize that the
experimental data of $\alpha$-decay energies are not involved in
the fit for the model parameters. We also note that the rms
deviations $\sigma (Q_\alpha)$ with the extended
Bethe-Weizs\"{a}cker (BW2) formula and the DZ mass models are
large, which might be due to the difficulty in determining the
right shell closures after $^{208}$Pb in these models where the
shell corrections of nuclei are described as a function of
valence-nucleon number. Fig. 3 shows the calculated $\alpha$-decay
energies of some super-heavy nuclei with the WS3 (solid curve) and
DZ28 models (open circles), respectively. The results from the WS3
model look much better than those from the DZ28 model, although
both models give similar rms deviation for known masses.

\begin{figure}
\includegraphics[angle=-0,width= 0.8 \textwidth]{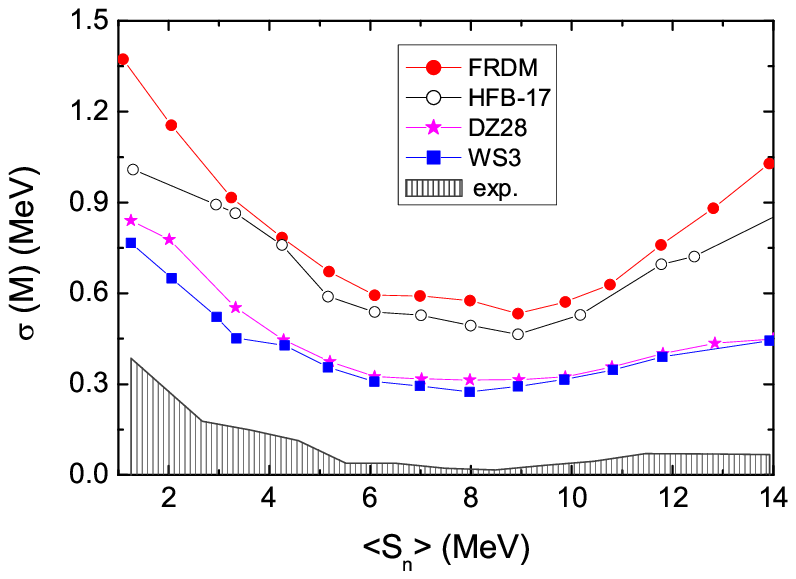}
 \caption{(Color online) rms deviation with respect to the masses as a function of average neutron-separation energy of nuclei.
 The shades present the average standard deviation errors of the measured masses \cite{Audi}.  }
\end{figure}

In Fig. 4 we show the deviations of the calculated total energies
of nuclei at their ground state with WS3 from the results of (a)
FRDM, (b) DZ28, (c) WS* and (d) HFB-17, respectively. The shades
denote the region with deviations smaller than 2 MeV. One sees
that the results from WS3, FRDM, DZ28 and WS* are relatively close
to each other, and the results from HFB-17 model  deviate
obviously from WS3 calculations for extremely neutron-rich nuclei.
We also note that the binding energies of super-heavy nuclei
around $N\sim 170$ and $Z\sim 100-110$ with DZ28 are significantly
larger by about 4 MeV than those with WS3, which leads to the
deviations in the calculation of the $\alpha$-decay energies (see
Fig. 3) between the two models. Because the predictions from
different models towards the neutron-drip line tend to diverge
\cite{Lun03}, it is necessary to analyze the systematic errors
from the various mass models for nuclei near the neutron drip
line. In Fig. 5 we show the rms deviations with respect to the
masses as a function of average neutron-separation energy $\langle
S_n \rangle$ of nuclei based on the masses from different mass
models. The shades present the average standard deviation errors
of the measured masses \cite{Audi}. One sees that the rms
deviations from the FRDM, HFB-17, DZ28 and WS3 increase
systematically for nuclei approaching the neutron drip line (i.e.,
the neutron separation energy $S_n\sim 0$),  which indicates that
the systematic errors for the neutron drip line nuclei from these
mass models expect to be much larger than those for stable nuclei.
The rms deviation $\sigma (M)$ from the FRDM and HFB-17 are about
1.4 and 1.0 MeV for nuclei with $S_n\approx 1$ MeV, respectively.
The corresponding results from WS3 and DZ28 are about 0.8 and 0.9
MeV, respectively. The experimental errors are about 0.4 MeV for
the masses of nuclei near the neutron drip line, and the tendency
of the errors for both experimental data and the model
calculations is similar with decreasing of $S_n$ of nuclei. For
nuclei approaching the proton drip line (with large value for
$S_n$), the rms deviations from DZ28 and WS3 are bout 0.45 MeV,
and the result from the FRDM reaches about 1 MeV. Fig. 5 indicates
that the model WS3 is relatively reliable for describing the
masses of nuclei far from the $\beta$-stability line.

\begin{figure}
\includegraphics[angle=-0,width= 0.7 \textwidth]{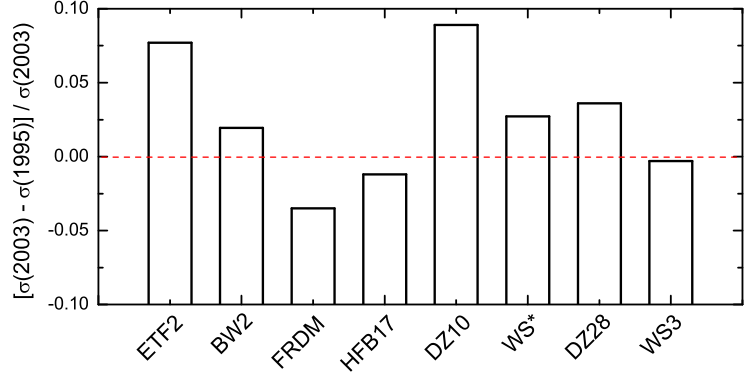}
 \caption{(Color online) A comparison of the predictive power of the various
models. The vertical axis denotes the quantity
$Y=\frac{\sigma(2003)-\sigma(1995)}{\sigma(2003)}$, as a measure
of the predictive power of the mass formulas fitted to the 1995
data. Here, $\sigma(1995)$  and $\sigma(2003)$  denote the rms
deviation to the masses of the AME1995 \cite{Audi95} and AME2003
\cite{Audi}, respectively.   See Table II for more information. }
\end{figure}

\begin{figure}
\includegraphics[angle=-0,width= 0.8\textwidth]{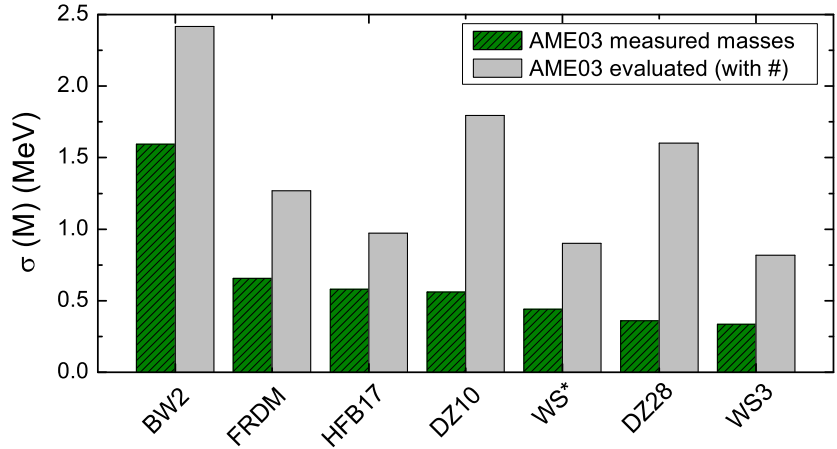}
 \caption{(Color online) A comparison of rms deviation with respect to the masses from AME2003 with the various
models. The bars with shades are for the 2149 measured masses and
the gray bars denote the rms deviations to the evaluated masses
(marked by $\#$) in AME2003. For the HFB-17 model, only the masses
of nuclei with $Z\le110$ are involved in the calculation. }
\end{figure}

Furthermore, to check the predictive power of various models, we
follow Ref.\cite{Lun03} and study the quantity
$Y=\frac{\sigma(2003)-\sigma(1995)}{\sigma(2003)}$, as a measure
of the predictive power of the mass formulas fitted to the 1995
data. Here, $\sigma(1995)$  and $\sigma(2003)$ denote the rms
deviation to the masses of the AME1995 \cite{Audi95} and AME2003
\cite{Audi}, respectively. Fig. 6 shows the values of $Y$ with
various mass model listed in Table II. Large positive value for
$Y$ indicates the deterioration for prediction of "new" masses in
2003 compilation. One can see from Fig. 6 that the predictive
power of FRDM, HFB-17 and WS3 seems to be relatively better than
those of DZ10 \cite{DZ10} and ETF2. In Fig. 7, we show a
comparison of the rms deviation with respect to the masses from
AME2003 with the various models. The bars with shades denote the
rms deviation to the 2149 measured masses and the gray bars denote
that to about 900 evaluated masses (marked by $\#$) in AME2003.
One sees that the predicted masses from WS3 are relatively close
to the evaluated masses in AME2003, with an rms deviation of 0.8
MeV (the evaluated masses in AME2003 are not involved in the fit
for the model parameters). The corresponding rms deviations from
the three models BW2, DZ10 and DZ28 in which one needs to
determine the magic numbers in the calculations are obviously
large than the others, like the large rms deviations with respect
to the $\alpha$-decay energies of super-heavy nuclei from these
three models [see the values of $\sigma(Q_\alpha)$ in Table II].
Figs. 2, 3, 5 and 6 demonstrate that the proposed model in this
work has a relatively high accuracy and good predictive power for
a description of the masses of the drip line nuclei and the
super-heavy nuclei.

\begin{center}
\textbf{IV. CONSTRAINTS ON MASS MODELS FROM LOCAL RELATIONS}
\end{center}

In this section, we perform a systematic test of the global
nuclear mass models by using the Garvey-Kelson relations and the
isobaric multiplet mass equation. We mainly compare the results
from the mass models mentioned previously. We first check the
Garvey-Kelson (GK) relations based on the obtained mass tables
from these mass models. It is known that the GK relation can not
be used to cross the $N=Z$ line. Therefore, we suitably select
five of the neighbors in GK relation for each nucleus to avoid
crossing the $N=Z$ line. We use the following GK relations:
\begin{eqnarray}
M_{\rm GK}(N,Z)&=&M(N+1,Z)+M(N,Z-1)+M(N+2,Z-2) \nonumber \\
      &-&M(N+1,Z-2)-M(N+2,Z-1) \; :  \; {\rm for} \; N\ge Z \; {\rm cases}
\end{eqnarray}
and
\begin{eqnarray}
M_{\rm GK}(N,Z)&=&M(N-1,Z)+M(N,Z+1)+M(N-2,Z+2) \nonumber \\
      &-&M(N-1,Z+2)-M(N-2,Z+1) \; :  \; {\rm for} \; N < Z \; {\rm cases}
\end{eqnarray}
Here, $M(N,Z)$ denotes the mass excess of a nucleus with neutron
number $N$ and charge number $Z$.  $\sigma ({\rm GK})$ in Table II
presents the rms deviation of the calculated masses by the
indicated mass models from the GK mass estimates according to
Eqs.(13) and (14). It is known that the ETF approach or the
Hartree-Fock-Bogoliubov (HFB) approach \cite{Dob} based on the
traditional Skyrme force without further improvements describe the
nuclear masses rather poorly, although the GK relations are
reproduced very well. With a series of improvements on the force,
the latest HFB-17 model \cite{HFB17} can reproduce the nuclear
masses with an accuracy of 581 keV. However the deviation of the
calculated masses from the GK mass estimates becomes relatively
large as mentioned in \cite{GK}. This indicates that more
constraints are still required for further testing the reliability
of the mass models, in addition to the GK relations. In other
words, fulfilling the GK relations is a necessary but not
sufficient condition to improve the quality of the model
prediction.

At the same time, we study the isobaric multiplet mass equation
(IMME) which is widely used to study the isospin-symmetry breaking
effect in nuclei. In the IMME, the binding energy  ($BE$) of a
nucleus is expressed as
\begin{eqnarray}
 BE(T,T_z)=a+b T_z+c T_z^2,
\end{eqnarray}
with $T_z= (N-Z)/2$. It is known that the IMME is valid if the
Coulomb interaction and charge-nonsymmetric parts of the
nucleon-nucleon interaction are weak relative to the strong
interaction, and it is generally assumed that the quadratic term
in the equation is adequate \cite{Lenzi,Ormand,LiBA}. The $b$
coefficients for light nuclei can be extracted from the measured
binding energies of pairs of mirror nuclei \cite{Ormand},
\begin{eqnarray}
b=\frac{BE(T=T_z)-BE(T=-T_z)}{2T}
\end{eqnarray}
with the isospin $T=|N-Z|/2$. Based on the liquid-drop model, one
can roughly estimate the value of the coefficient $b \sim  a_c
A^{2/3}$. Here, $a_c$ denotes the Coulomb energy coefficient in
the liquid-drop model. In \cite{Ormand}, Ormand proposed a formula
\begin{eqnarray}
 b_{\rm fit}=0.710A^{2/3}-0.946
\end{eqnarray}
for the $b$ coefficients by fitting the experimental data of $A <
60$. In Fig. 8 (a), we show the extracted $b$ coefficients of
nuclei in the IMME from mirror nuclei with $A\le75$ according to
Eq.(16). Here, the experimental binding energies of mirror nuclei
are taken from AME2003 \cite{Audi} and the four very recently
measured masses for $^{63}$Ge, $^{65}$As, $^{67}$Se, and $^{71}$Kr
in Ref. \cite{Xu}. It seems to be that the $b$ coefficients are
roughly constant for a given $A$. The solid curve denotes the
results of the $b_{\rm fit}$ from Eq.(17). For intermediate and
heavy nuclei, the $b$ coefficients could be extracted by fitting
the measured binding energies of a series of isobaric nuclei with
parabolas. Here, we remove the shell corrections  \cite{Wang10}
and the Wigner energies (about $47|I|$ MeV) \cite{Sat} of nuclei
from the measured binding energies of nuclei before performing a
parabolic fitting in order to eliminate the oscillations in the
results. The solid (open) circles denote the extracted results for
a series of isobaric even-$A$ (odd-$A$) nuclei, with $95\%$
confidence interval given by the error bars. One can see that the
trend of the $b$ coefficients with $A$ can be reasonably well
described by the empirical formula $b_{\rm fit}$, which indicates
that the IMME is valid for both light and heavy nuclei in general.
The residual oscillations in the $b$ coefficients might come from
the deformations of nuclei.

\begin{figure}
\includegraphics[angle=-0,width= 0.9\textwidth]{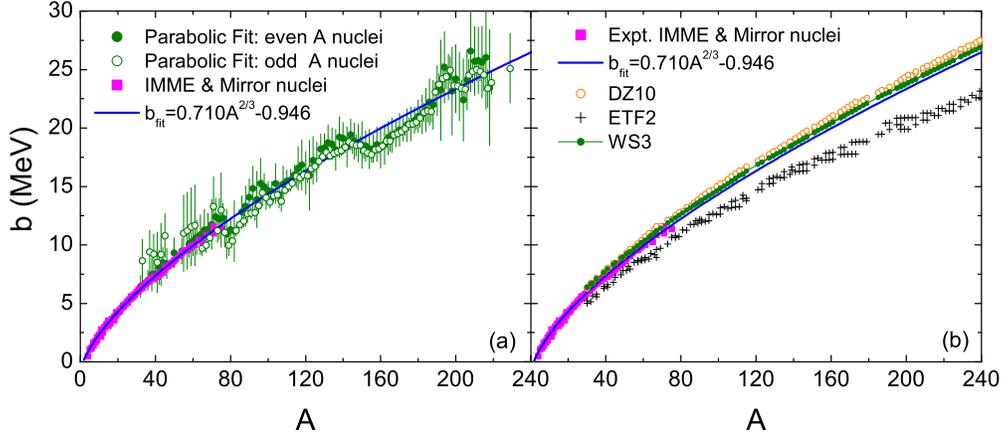}
 \caption{(Color online) (a) $b$ coefficients of nuclei in the IMME as a function of mass number. The solid squares and the solid curve denote the extracted results from mirror nuclei with $A\le75$ according to
 Eq.(16) and those from the empirical formula \cite{Ormand} $b_{\rm fit}=0.710A^{2/3}-0.946$,
 respectively. The solid (open) circles
 denote the extracted results by fitting the measured
 binding energies with parabolas for a series of isobaric even-$A$ (odd-$A$) nuclei. See text for details. (b) The same as (a), but with the results from some mass models for comparison.
The crosses, the open circles and the filled circles
 denote the results for 179 neutron drip line nuclei from ETF2, DZ10 and WS3 calculations, respectively.}
\end{figure}

According to the macroscopic-microscopic mass model, the binding
energy of a nucleus can be expressed as
\begin{eqnarray}
 BE(A,T_z)=\tilde{a} +\tilde{b} \,  T_z+\tilde{c}\,  T_z^2 + B_{\rm sh}+B_{\rm wig}+B_{\rm
 def}+...
\end{eqnarray}
Here, the mass-dependent coefficients $\tilde{a} $, $\tilde{b}$
and $\tilde{c} $ can be directly obtained from the liquid-drop
formula, and $\tilde{b}\sim a_c A^{2/3}$. $B_{\rm sh}$, $B_{\rm
wig}$ and $ B_{\rm def}$ denote the shell correction, the Wigner
energy and the deformation energy of a nucleus, respectively. From
Eqs.$(15) \sim (18)$ and Fig. 8(a), we learn that: 1) the shell
correction difference between a nucleus and its mirror nucleus is
small due to the isospin symmetry in nuclear physics as mentioned
in our previous work \cite{Wang10}; 2) the Wigner energy of a
nucleus equals generally to that of its mirror nucleus  since it
is usually expressed  as a function of $|I|$; and 3) the
deformation energy of a nucleus might be close to the value of its
mirror nucleus. The mirror nuclei effects from the isospin
symmetry could give a strong constraint on nuclear mass models.

The crosses, the open circles and the filled circles in Fig. 8(b)
denote the calculated results with Eq.(16) for 179 neutron drip
line nuclei ($Z\ge 8$ and $N\le200$) from ETF2, DZ10 and WS3
calculations, respectively. The results from WS3 are relatively
close to the values of $b_{\rm fit}$ (with an rms deviation of 449
keV), which indicates that the IMME with quadratic form is
generally satisfied in this model. For extremely neutron-rich
nuclei, the results (crosses) from the traditional Skyrme EDF
obviously deviate from the values of $b_{\rm fit}$. From  Fig.
9(a), we note that the change of $b_{\rm fit}-b$ with the isospin
$T$ for the 179 neutron drip line nuclei is quite different with
different models. For the Skyrme EDF calculations the difference
$b_{\rm fit}-b$ increases soon with the isospin $T$ (we draw the
same conclusion with Skyrme HFB codes HFBRAD \cite{HFBRAD} and
HFBTHO \cite{HFBTHO}). Without special consideration for the
isospin-symmetry restoration in the traditional Skyrme EDF, the
calculated surface diffuseness of protons becomes abnormally large
in extremely proton-rich nuclei to reduce the huge Coulomb
repulsion, which causes the large deviations from $b_{\rm fit}$.
These investigations show that the exploration of the IMME in the
neutron-rich nuclei might provide us with very useful information
for testing the mass models.

Closely relating to the IMME, the Coulomb energies (direct term)
of nuclei are investigated simultaneously. Fig. 9(b) shows the
Coulomb energy coefficient defined as $a_c=E_{C} A^{1/3} Z^{-2}$
with different models as a function of $(N-Z)/2$ for Pb isotopes.
Here, $E_C$ denotes the calculated Coulomb energy of a nucleus
with a certain mass model. The open circles and the solid line
denote the values of $a_c$ with DZ10 and WS3, respectively. The
squares and the crosses denote the results from the HFBTHO
\cite{HFBTHO} and the ETF2 \cite{Lium} approach with the Skyrme
force SkP \cite{SkP}, respectively. Similar to Fig. 9(a),
 the trends are quite different with different models. The
coefficient $a_c$ in this work is a constant. In DZ mass models,
the isospin effect in the Coulomb energy calculation is involved
by the isospin-dependent charge radius $R_c\propto
A^{1/3}(1-0.25I^2)$, which causes the reduction of the effective
coefficient $a_c$ for neutron-rich nuclei (see the open circles).
On the contrary, the values of $a_c$ obviously increase with
increasing neutron number in the Skyrme energy-density functional
calculations. It is because for extremely proton-rich nuclei the
surface diffuseness of protons increases sharply to reduce the
huge Coulomb repulsion and causes the reduction of $a_c$
consequently.  Comparing Fig. 9(a) to (b), one finds that the
change of $b_{\rm fit}-b$ with the isospin $T$ is roughly
consistent with that of the Coulomb energy coefficient $a_c$. The
Coulomb energy coefficient might be used as a sensitive quantity
to study the mass models and the isobaric multiplet mass equation.

\begin{figure}
\includegraphics[angle=-0,width= 1\textwidth]{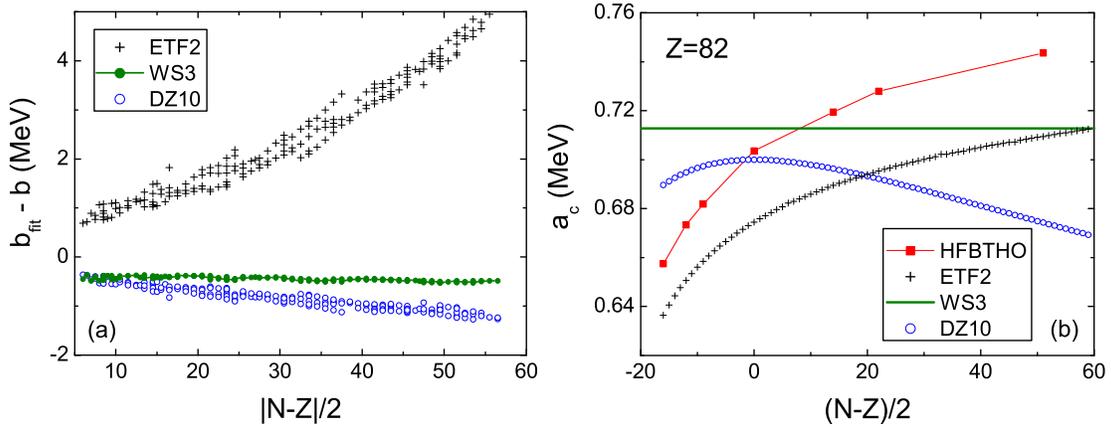}
 \caption{(Color online) (a) Deviations of the calculated $b$ coefficients with DZ10, ETF2 and
WS3 models from $b_{\rm fit}=0.710A^{2/3}-0.946$ \cite{Ormand}, as
a function of isospin. (b) Coulomb energy coefficient $a_c=E_{C}
A^{1/3} Z^{-2}$ obtained with different models as a function of
$(N-Z)/2$. }
\end{figure}

\begin{figure}
\includegraphics[angle=-0,width= 0.7\textwidth]{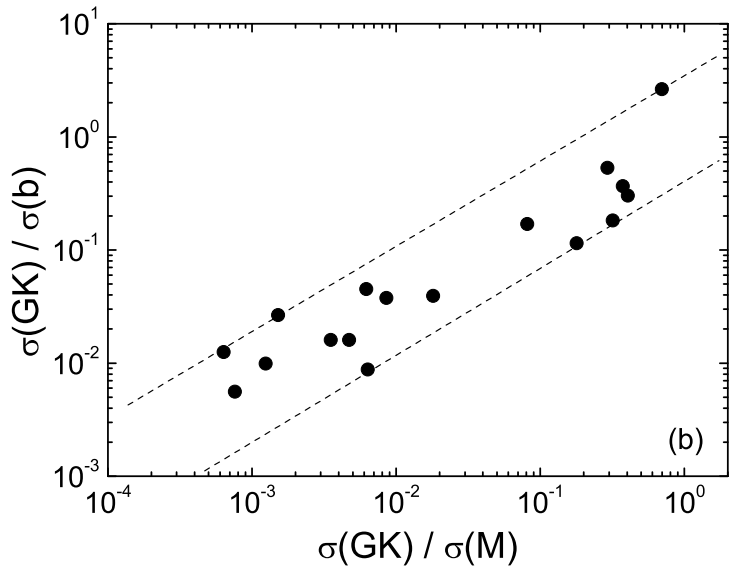}
 \caption{ Values of $\sigma$(GK)/$\sigma (b)$ as a
function of $\sigma$(GK)/$\sigma$(M) from 17 different models
(solid circles). Dashed lines are to guide the eyes.  }
\end{figure}

In addition, for exploring the correlation among the IMME, the GK
relation and the accuracy of the nuclear mass models, we
systematically study the rms deviations of calculated $b$
coefficients with Eq.(16) and the nuclear masses by 17 nuclear
mass models, including different versions of liquid drop models
\cite{Moll95,BW,Pear01,Isa}, the Duflo-Zuker mass models (DZ10 and
DZ31 \cite{DZ10}), the ETF2 approach with different Skyrme forces,
the WS, WS* and WS3 models. We use $\sigma (b)$ for describing the
rms deviation of the calculated $b$ coefficients by a certain mass
model from $b_{\rm fit}$ for the 179 neutron drip line nuclei. The
rms deviations $\sigma (b)$ obtained from ETF2, DZ10, DZ31
\cite{DZ10} and WS3 are 2734, 870, 630 and 449 keV, respectively.
The values of $\sigma (M)$ and $\sigma$(GK) from some mass models
are presented in Table II previously. In Fig. 10, we show the
values of $\sigma$(GK)/$\sigma (b)$ as a function of
$\sigma$(GK)/$\sigma$(M).  Each of solid circles represents the
calculated result from a nuclear mass model and all of circles are
located in between two dashed lines. A general tendency of
$\sigma$(GK)/$\sigma (b)$ increasing with $\sigma$(GK)/$\sigma
(M)$ is clearly shown, which exhibits a strong correlation between
the quadratic form of the IMME and the accuracy of the mass model
when the GK relations are reproduced reasonably well. In other
words, reducing the deviation from the IMME could be a way to
improve the accuracy of mass predictions if the $\sigma$(GK) $\sim
100$ keV. Fulfilling both the IMME and the GK relations seems to
be two necessary conditions to improve the reliability of mass
calculations for known nuclei and the extremely neutron-rich
nuclei.

\newpage

\begin{center}
\textbf{IV. SUMMARY}
\end{center}

We proposed an extension of our earlier global nuclear mass model
which significantly improves its ability to describe nuclear
masses across the periodic table and puts it at least on a par
with the very best empirical mass models on the market. With some
additional correction terms, including mirror nuclei effect due to
the isospin symmetry, Wigner-like effect from the symmetry between
valance-neutron and valance-proton, corrections for pairing
effects and those from triaxial deformation of nuclei, the rms
deviations from 2149 measured masses and 1988 neutron separation
energies are significantly reduced to 336 and 286 keV,
respectively. As a test of the extrapolation of the mass model,
the $\alpha$-decay energies of 46 super-heavy nuclei have been
systematically studied. The rms deviation with the proposed model
reaches 248 keV, much smaller than the result 936 keV from the
DZ28 model. Furthermore, through studying the rms deviations to
the masses of nuclei approaching the drip lines and the predictive
power of the mass formulas fitted to the AME1995 data, we find
that the results from the proposed model are satisfactory. In
addition, with a systematic study of 17 global nuclear mass
models, we find that the quadratic form of the IMME is closely
related to the accuracy of nuclear mass calculations when the
Garvey-Kelson relations are reproduced reasonably well. Fulfilling
both the IMME and the Garvey-Kelson relations seems to be two
necessary conditions to improve the quality of the model
prediction. Furthermore, the $\alpha$-decay energies of
super-heavy nuclei should be used as an additional constraint on
the global nuclear mass models.

\begin{center}
\textbf{ACKNOWLEDGEMENTS}
\end{center}

We are grateful to Professors Z. X. Li, W. Scheid, E. G. Zhao and
S. G. Zhou for valuable suggestions. This work was supported by
National Natural Science Foundation of China, Nos 10875031,
10847004, 11005022 and 10979024. The obtained mass table with the
proposed formula is available at
http://www.imqmd.com/wangning/WS3.6.zip.

\end{document}